\documentclass{article}
\usepackage{a4wide}
\usepackage{amssymb}
\usepackage{amsmath}
\usepackage{amsfonts}
\usepackage{amsthm}
\usepackage{epsfig}

\usepackage[section]{algorithm}
\usepackage{algorithmic}
\usepackage{url}
\newcommand{\GO}{{\cal O}}
\newtheorem{thm}{Theorem}[section]
\newtheorem{lem}[thm]{Lemma}

\begin{document}

\author{Jean-Guillaume Dumas\footnote{
   Universit\'e de Grenoble, laboratoire de mod\'elisation
   et calcul, LMC-IMAG BP 53 X, 51 avenue des math\'ematiques,
   38041 Grenoble, France.
   \texttt{\{Jean-Guillaume.Dumas,~Clement.Pernet\}@imag.fr}
   .}~~and Cl\'ement Pernet\footnotemark[1] and Zhendong
   Wan\footnote{Dept. of Computer and Inf. Science, University of
   Delaware, Newark, DE 19716, USA. \texttt{Wan@cis.udel.edu}.}}
\date{\today}

%

\title{Efficient Computation of the Characteristic Polynomial}



\date{\today}

\maketitle

\begin{abstract}
This article deals with the computation of the characteristic
polynomial of dense matrices over small finite fields and over the
integers. We first present two algorithms for the finite fields: one
is based on Krylov iterates and Gaussian elimination. We compare it to
an improvement of the second algorithm of Keller-Gehrig. 
Then we show that a generalization of Keller-Gehrig's third algorithm
could improve both complexity and computational time.
We use these results as a basis for the computation of the characteristic
polynomial of integer matrices. We first use early termination and
Chinese remaindering for dense matrices.
Then a probabilistic approach, based on integer minimal polynomial and
Hensel factorization, is particularly well suited to sparse and/or
structured matrices.

\end{abstract}




\section{Introduction}

Computing the characteristic polynomial of an integer matrix is a classical mathematical problem. 
It is closely related to the computation of the Frobenius normal form which can be used to test
two matrices for similarity. Although the Frobenius normal form contains more information on 
the matrix than the characteristic polynomial, most algorithms to compute it are based on 
computations of characteristic polynomial (see for example
\cite[\S 9.7]{Storjohann:2000:thesis} ).

Using classic matrix multiplication, the algebraic time complexity of
the computation of the characteristic polynomial is nowadays optimal. Indeed,
many algorithms  have a $\GO(n^3)$  algebraic time complexity ( to our
knowledge the older one is due to Danilevski, described in \cite[\S
  24]{HouseHolder:1964:TOMINA}). The fact that the computation of
the determinant is proven to be as hard as matrix multiplication \cite{BaurStrassen:83:deriv}
ensures this optimality. But with fast matrix arithmetic
($\GO(n^\omega)$ with $2\leq\omega<3$), the best asymptotic time
complexity is $\GO(n^\omega \text{log}n)$, given by Keller-Gehrig's
branching algorithm \cite{Keller-Gehrig:1985}. Now the third algorithm of
Keller-Gehrig has a $\GO(n^\omega)$ algebraic time complexity but only works
for generic matrices.

In this article we focus on the practicability of such algorithms
applied on matrices over a finite field. Therefore we used the techniques
developped in \cite{Dumas:2002:FFLAS,Dumas:2004:FFPACK}, for efficient basic
linear algebra operations over a finite field. We propose a new
$\GO(n^3)$ algorithm designed to take benefit of the block matrix
operations; improve Keller-Gehrig's branching algorithm and compare
these two algorithms. Then we focus on Keller-Gehrig's third algorithm
and prove that its generalization is not only of theoretical interest
but is also promising in practice.

As an application, we show that these results directly lead to an
efficient computation of the characteristic polynomial of integer
matrices using chinese remaindering and an early termination criterion
adaptated from \cite{DumasSaundersVillard:2001:JSC}. This basic
application outperforms the best existing softwares on many cases. 
Now better algorithms exist for the integer case, and can be more
efficients with sparse or structured matrices. 
Therefore, we also propose a probabilistic algorithm using a black-box
computation of the minimal polynomial and our finite field algorithm.
This can be viewed as a simplified version of the algorithm
described in \cite{Storjohann:2000:Frob} and \cite[\S 7.2]{KaltofenVillard:2004:det}. Its efficiency
in practice is also very promising.


\section{Krylov's approach}
Among the different techniques to compute the characteristic
polynomial over a field, many of them rely on the Krylov approach. A
description of them can be found in \cite{HouseHolder:1964:TOMINA}.
They are based on the following fact: the minimal linear dependance
relation between the Krylov iterates of a vector $v$ (i.e. the
sequence $(A^iv)_i$ gives the minimal polynomial $P_{A,v}^{min}$ of this sequence, and a
divisor of the minimal polynomial of $A$. Moreover, if $X$ is the
matrix formed by the first independent column vectors of this
sequence, we have the relation 
$$AX=XC_{P_{A,v}^{min}}$$
where $C_{P_{A,v}^{min}}$ is the companion matrix associated to
$P_{A,v}^{min}$.

\subsection{Minimal polynomial} \label{sec:minpoly}

We give here a new algorithm to compute the minimal polynomial of the sequence
of the Krylov's iterates of a vector $v$ and a matrix $A$. This is the
monic polynomial $P_{A,v}^{min}$ of least degree such that $P(A).v = 0$.
We firstly presented it in \cite{PernetWan:2003:Charp, Pernet:2003:dea} and it was
simultaneously published in \cite[Algorithm 2.14]{LomAbd:04:Methodes}.

\begin{figure}[htbp]
\includegraphics[width=\textwidth]{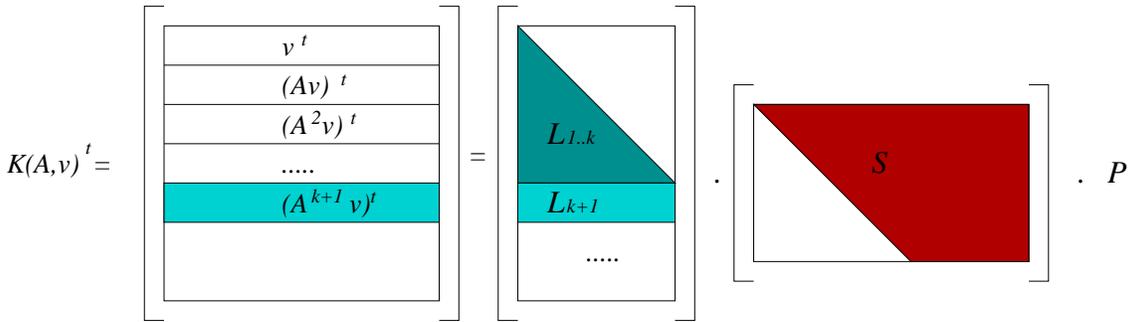}
\caption{principle of the computation of $P_{A,v}^{min}$}
\label{fig:minpoly}
\end{figure}

The idea is to compute the $n \times n$ matrix $K_{A,v}$ (we call it
  Krylov's matrix), whose $i$th column is the vector $A^iu$, and to perform
  an elimination on it. More precisely, one computes the LSP
  factorization of $K_{A,v}^t$ (see \cite{Ibarra:1982:LSP} for a description of the LSP
  factorization). Let $k$ be the degree of
  $P_{A,v}^{min}$. This means that the first $k$ columns of $K_{A,v}$ 
  are linearly independent, and the $n-k$ following ones are linearly dependent
  with the first $k$ ones. Therefore $S$ is triangular with its last $n-k$ rows
  equals to $0$. Thus, the LSP factorization of  $K_{A,v}^t$ can
  be viewed as in figure \ref{fig:minpoly}.

Now the trick is to notice that the vector $m=L_{k+1}L_{1 \dots
  k}^{-1}$ gives the opposites of the coefficients of $P_{A,v}^{min}$.
Indeed, let us define $X=K_{A,v}^t$
\[ 
X_{1\dots n,k+1} = (A^kv)^t = \sum_{i=0}^{k-1}{m_i(A^iv)^t} = m \cdot
X_{1 \dots n,1\dots k}
\]
where $P_{A,v}^{min}(X) =  X^k -m_kX^{k-1} - \dots - m_1X- m_0 $.

Thus
\[
L_{k+1}SP = m \cdot L_{1\dots k}SP
\] 
And finally $m = L_{k+1}.L_{1\dots k}^{-1}$

The algorithm is then straightforward:

\begin{algorithm}
\caption{\texttt{MinPoly}~: Minimal Polynomial of $A$ and $v$} \label{alg:minpoly}
\begin{algorithmic} [1]
\REQUIRE{$A$ a $n\times n$ matrix and $v$ a vector over a field}
\ENSURE{$P_\text{min}^{A,v}(X)$ the minimal polynomial of the sequence
of vectors $(A^iv)_i$}
\STATE $K_{1 \dots n,1} = v$ 
\FOR{$i=1$ to $\text{log}_2(n)$}
 \STATE $K_{1 \dots n,2^i \dots 2^{i+1}-1} = A^{2^{i-1}}K_{1 \dots n,1 \dots 2^i-1}$
\ENDFOR 
 \STATE $(L,S,P) = \mbox{LSP}(K^t), k = \mbox{rank}(K)$
 \STATE $m = L_{k+1}.L_{1\dots k}^{-1}$
 \STATE return $P_\text{min}^{A,v}(X)=X^k +\sum_{i=0}^{k-1}{m_iX^i}$
\end{algorithmic}
\end{algorithm}

The dominant operation in this algorithm is the computation of $K$, in
$log_2 n$ matrix multiplications, i.e. in $ \GO (n^\omega log n)$ algebraic
operations. The LSP factorization requires $ \GO (n^\omega)$
operations and the triangular system resolution, $ \GO (n^2)$.
The algebraic time complexity of this algorithm is thus $ \GO
(n^\omega log n)$.

When using classical matrix multiplications (assuming $\omega = 3$),
it is preferable to compute the Krylov matrix $K$ by $k$ successive
matrix vector products. The number of field operations is then $\GO
(n^3)$.

It is also possible to merge the creation of the Krylov matrix and its
LSP factorization so as to avoid the computation of the last $n-k$
Krylov iterates with an early termination approach. This reduces the
time complexity to $\GO (n^\omega log(k))$ for fast matrix arithmetic,
and $\GO(n^2k)$ for classic matrix arithmetic.

Note that choosing $v$ randomly makes the algorithm Monte-Carlo for
the computation of the minimal polynomial of A.



\subsection{LU-Krylov algorithm} \label{sec:luk}

We present here an algorithm, using the previous computation of the
minimal polynomial of the sequence $(A^iv)_i$ to compute the
characteristic polynomial of $A$. 
The previous algorithm produces the $k$ first independent Krylov
iterates of $v$. They can be viewed as a basis of an invariant
subspace under the action of $A$, and if $P_{A,v}^{min} = P_A^{min}$,
this subspace is the first invariant subspace of $A$. The idea is to
make use of the elimination performed on this basis to compute a basis
of its supplementary subspace. Then a recursive call on this second
basis will decompose this subspace into a series of invariant
subspaces generated by one vector.

The algorithm is the following, where $k$, $P$, and $S$ come from the
notation of algorithm \ref{alg:minpoly}.

\begin{algorithm}
\caption{\texttt{LUK}~: LU-Krylov algorithm} \label{alg:luk}
\begin{algorithmic} [1]
\REQUIRE{$A$ a $n\times n$ matrix over a field}
\ENSURE{$P_\text{char}^A(X)$ the characteristic polynomial of $A$}
\STATE Pick a random vector $v$
\STATE $P_\text{min}^{A,v}(X)=MinPoly(A,v)$ of degree $k$\\
\COMMENT {$X=\left[\begin{array}{c}L_1\\L_2\end{array}\right]
[S_1|S_2] P$ is computed}
\IF { $(k = n)$ }
 \STATE return $P^A_\text{char} = P^{A,v}_\text{min}$
\ELSE 
 \STATE $A' = PA^TP^T =
 \left[\begin{array}{cc}A'_{11}&A'_{12}\\A'_{21}&A'_{22}\end{array}\right]$
 where $A'_{11}$ is $k\times k$.
 \STATE 
 $P^{A'_{22}-A'_{21}S_1^{-1}S_2}_\text{char}(X) = LUK(A'_{22}-A'_{21}S_1^{-1}S_2)$
 \STATE return $P^A_\text{char}(X) = P^{A,v}_\text{min}(X) \times P^{A'_{22}-A'_{21}S_1^{-1}S_2}_\text{char}(X) $
\ENDIF
\end{algorithmic}
\end{algorithm}

\begin{thm}
The algorithm LU-Krylov computes the characteristic polynomial of an
$n \times n$ matrix $A$ in $\GO (n^3)$ field operations.
\end{thm}

\begin{proof}
Let us use the following notations
\[
X=\left[\begin{array}{c}L_1\\L_2\end{array}\right] [S_1|S_2] P
\]

As we already mentioned, the first $k$ rows of $X$ ($X_{1 \dots k,
  1 \dots n}$) form a basis of the invariant subspace generated by
  $v$. Moreover we have
$$
X_{1..k}A^T=C_{P^{A,v}_\text{min}}^T X_{1..k}
$$
Indeed
\[
\forall i<k \ X_i A^T = \left(A^{i-1}v\right)^TA^T = \left(A^iv\right)^T=X_{i+1}
\]
and
$$
X_k A^T = \left(A^{k-1}v\right)^TA^T = \left(A^kv\right)^T = \sum_{i=0}^{k-1}{m_i\left(A^iv\right)^T}
$$

The idea is now to complete this basis into a basis of the whole space.
Viewed as a matrix, this basis form the $n \times n$ invertible matrix
$\overline{X}$. It is defined as follows:

\[
\overline X = 
\underbrace{\left[\begin{array}{cc} L_1 & 0\\0&I_{n-k}\end{array}\right]}_{\overline L}
\underbrace{\left[\begin{array}{cc} S_1 & S_2\\0&I_{n-k}\end{array}\right]}_{\overline S}
 P = 
\left[\begin{array}{c} 
    X_{1 \dots k,1 \dots n} \\ 
    \left[\begin{array}{cc} 0&I_{n-k}\end{array}\right] P 
\end{array}\right]
\]

Let us compute
\begin{eqnarray*}
\overline X A^T \overline X^{-1} &=& 
\left[
\begin{array}{c|c} 
\hspace{8mm} C^T \hspace{8mm}  & 0\\
\hline
\multicolumn{2}{c}
{
\left[\begin{array}{cc}0&I_{n-k} \end{array} \right] PA^TP^T
     {\overline S}^{-1}{\overline L}^{-1}
}
\end{array} 
\right]
\\&=&
\left[
\begin{array}{c|c} 
\hspace{6mm} C^T \hspace{6mm}& 0\\
\hline
\multicolumn{2}{c}
{
\left[\begin{array}{cc} A'_{21}&A'_{22} \end{array} \right]
     {\overline S}^{-1}{\overline L}^{-1}
}
\end{array}
\right]
\\ &=&
\left[
\begin{array}{c|c} 
C^T & 0\\
\hline
Y & X_2
\end{array}
\right]
\end{eqnarray*}

with
\[
X_2= A'_{22}-A'_{21}S_1^{-1}S_2
\]

By a similarity transformation, we thus have reduced $A$ to a block
triangular matrix. Then the characteristic polynomial of $A$ is the
product of the characteristic polynomial of these two diagonal blocks:
$$ P_\text{char}^A =  P^{A,v}_\text{min} \times
P^{A'_{22}-A'_{21}S_1^{-1}S_2}_\text{char}
$$

Now for the time complexity, we will denote by $T_{\tt LUK}(n)$ the
number of field operations for this algorithm applied on a $n \times
n$ matrix, by $T_{\tt minpoly}(n,k)$ the cost of the algorithm
\ref{alg:minpoly} applied on a $n \times n$ matrix having a degree $k$
minimal polynomial, by $T_{\tt LSP}(m,n)$ the cost of the LSP factorization of
a $m \times n$ matrix, by $T_{\tt trsm}(m,n)$ the cost of the
simultaneous resolution of $m$ triangular systems of dimension $n$,
and by $T_{\tt MM}(m,k,n)$ the cost of the multiplication of a $m
\times k$ matrix by a $k \times n$ matrix.

The values of $T_{\tt LSP}$ and
$T_{\tt trsm}$ can be found in \cite{Dumas:2004:FFPACK}.
Then, using classical matrix arithmetic, we have:
\begin{eqnarray*}
T_{\tt LUK}(n) & = &  T_{\tt minpoly}(n,k) + T_{\tt LSP}(k,n) + T_{\tt
trsm}(n-k,k) \\
               &   & + T_{\tt mm}(n-k,k,n-k) + T_{\tt LUK}(n-l)\\
               & = & \GO(n^2k+k^2n+k^2(n-k)+k(n-k)^2) \\
               &   & + T_{\tt LUK}(n-k)\\
               & = & \GO(\sum_i{n^2k_i+k_i^2n}) \\
               & = & \GO(n^3)
\end{eqnarray*}
The latter being true since $\sum_i k_i = n$ and $\sum_i{k_i^2} \leq n^2$.

\end{proof}

Note that when using fast matrix arithmetic, it is no longer
possible to sum the $log(k_i)$ into $log(n)$ or the $k_i^{\omega-2}n^2$
into $n^\omega$, so this prevents us from getting the best known time
complexity of $n^\omega log(n)$ with this algorithm. 
We will now focus on the second algorithm of Keller-Gehrig achieving
this best known time complexity. 

\subsection{Improving Keller-Gehrig's branching algorithm} \label{sec:kgb}

In \cite{Keller-Gehrig:1985}, Keller-Gehrig presents a so called
branching algorithm, computing the characteristic polynomial of a $n
\times n$ matrix over a field $K$ in the best known time complexity of
$n^\omega log(n)$ field operations.

The idea is to compute the Krylov iterates of a several vectors at the
same time. More precisely, the algorithm computes a sequence of $n
\times n$ matrices $(V_i)_i$ whose columns are the Krylov's iterates of vectors of
the canonical basis. $U_0$ is the identity matrix (every vector of the
canonical basis is present). At the $i$-th iteration, the algorithm computes
the following $2^i$ Krylov's iterates of the remaining vectors. Then
a Gaussian elimination determines the linear dependencies between them
so as to form $V_{i+1}$ by picking the $n$ linearly independent
vectors.
The algorithm ends when each $V_i$ is invariant under the
action of $A$.  Then the matrix $V^{-1}AV$ is block diagonal with
companion blocks on the diagonal. The polynomials of these blocks are
the minimal polynomials of the sequence of Krylov's iterates, and the characteristic polynomial is
the product of the polynomials associated to these companion blocks.

The  linear dependencies removal is performed by a step-form elimination
algorithm defined by Keller-Gehrig. Its formulation is rather
sophisticated, and we propose to replace it by the column reduced form
algorithm (algorithm \ref{alg:CRF}) using the more standard LQUP
\begin{algorithm}
\caption{\texttt{ColReducedForm}}\label{alg:CRF}
\begin{algorithmic}[1]
\REQUIRE{$A$ a $m \times n$ matrix of rank $r$ ($m,n\geq r$) over a field}
\ENSURE{$A'$ a $m \times r$ matrix formed by r linearly independent
columns of $A$}
\STATE $(L,Q,U,P,r)=\text{LQUP}(A^T)$ ($r=rank(A)$)
\STATE \textbf{return} $([I_r 0](Q^TA^T))^T$
\end{algorithmic}
\end{algorithm}
factorization (described in \cite{Ibarra:1982:LSP}).
More precisely, 
the step form elimination of Keller-Gehrig, the LQUP factorization of
Ibarra \& Al. and the echelon elimination (see
e.g. \cite{Storjohann:2000:thesis}) are equivalent and can be
used to determine the linear dependencies in a set of vectors.

\begin{algorithm}[phtb]
\caption{\texttt{KGB}: Keller-Gehrig Branching algorithm}\label{alg:kg2}
\begin{algorithmic}[1]
\REQUIRE{$A$ a $n \times n$ matrix over a field}
\ENSURE{$P_\text{char}^A(X)$ the characteristic polynomial of $A$}
\STATE $i=0$
\STATE $V_0 = I_n = (V_{0,1},V_{0,2},\dots,V_{0,n})$
\STATE $ B=A $
\WHILE{($\exists k, V_k$ has $2^i$ columns)}
 \FORALL j
  \IF{ ( $V_{i,j}$ has strictly less than $2^i$ columns )}
  \STATE $W_j = V_{i,j}$
  \ELSE
  \STATE $W_j = \left[ V_{i,j} |BV_{i,j}\right]$
  \ENDIF
 \ENDFOR
 \STATE $W = (W_j)_j$
 \STATE $V_{i+1} = \text{ColReducedForm}( W )$\\
 \COMMENT{$V_{i+1,j}$ are the remaining vectors of $W_j$ in $V_{i+1}$}
 \STATE $B = B \times B$
 \STATE $i=i+1$
\ENDWHILE
\FORALL{$j$}
 \STATE compute $P_j$ the minimal polynomial of the sequence of
 vectors of $V_{i-1,j}$, using algorithm \ref{alg:minpoly}
\ENDFOR
\STATE \textbf{return} $\Pi_j{P_j}$
\end{algorithmic}
\end{algorithm}

Our second improvement is to apply the idea of algorithm \ref{alg:minpoly} to compute
polynomials associated to each companion block, instead of computing
$V^{-1}AV$. The Krylov's iterates are already computed, and the last
call to \texttt{ColReducedForm} performed the elimination on it, so there
only remains to solve the triangular systems so as to get the
coefficients of each polynomial.

Algorithm \ref{alg:kg2} sums up these modifications. 
The operations in the \texttt{while} loop have a $\GO(n^\omega)$
algebraic time complexity. This loop is executed at most
$\text{log}(n)$ times and the algebraic time complexity of the
algorithm is therefore $\GO(n^\omega\text{log}(n))$. More precisely it
is $\GO(n^\omega\text{log}(k_\text{max}))$ where $k_\text{max}$ is the
degree of the largest invariant factor.

\subsection{Experimental comparisons}\label{sec:krylovexp}

\begin{figure}[htbp]  
\includegraphics[width=\textwidth]{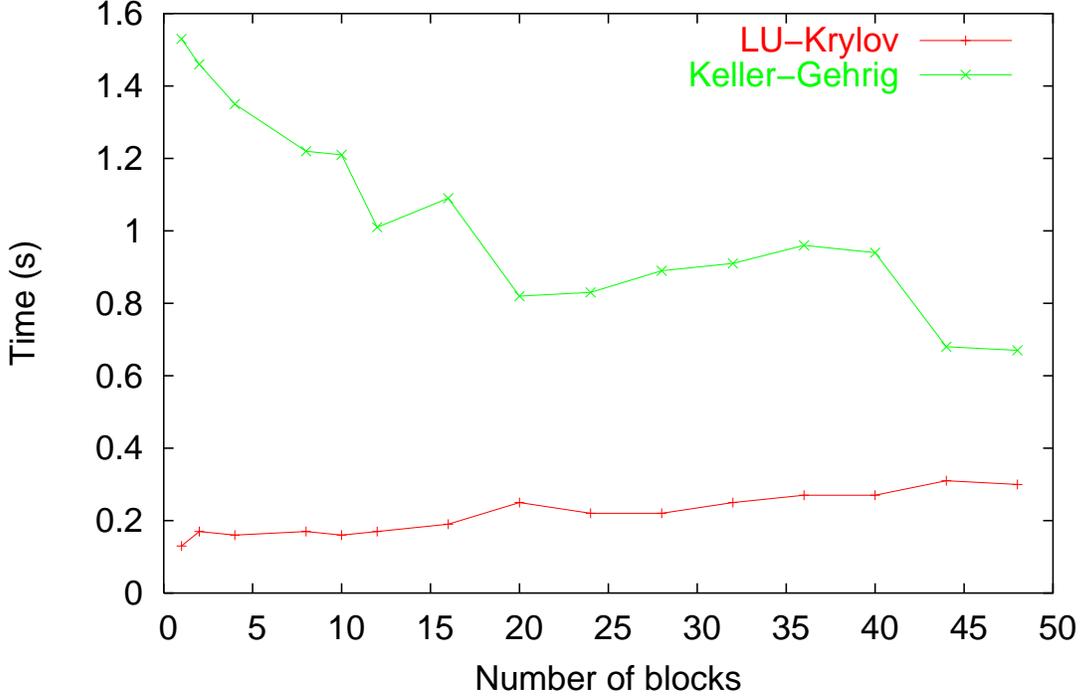} 
\caption{LU-Krylov vs. KGB} \label{fig:kgvluk}
\end{figure}

To implement these two algorithms, we used a finite field
representation over double size floating points: {\tt modular<double>} 
(see \cite{Dumas:2004:FFPACK}) and the efficient routines for finite field
linear algebra {\tt FFLAS-FFPACK} presented in
\cite{Dumas:2004:FFPACK,Dumas:2002:FFLAS}. The following experiments
used a classic matrix
arithmetic.
We ran them on a series of matrices of order $300$ which  Frobenius
normal forms had different number of diagonal companion blocks.
Figure \ref{fig:kgvluk} shows the computational time
on a Pentium IV 2.4Ghz with 512Mb of RAM.

It appears that \texttt{LU-Krylov} is faster than \texttt{KGB} on
every matrices. This is due to the extra $\text{log}(n)$ factor in the
time complexity of the latter.
One can note that the computational time of \texttt{KGB} is decreasing
with the number of blocks. This is due to the fact that the
$\text{log}(n)$ is in fact $\text{log}(k_\text{max}$ where
$k_\text{max}$ is the size of the largest block. This factor is decreasing
when the number of blocks increases. Conversely, \texttt{LU-Krylov}
computational time is almost constant. It slightly increases, due to
the increasing number of rectangular matrix operations. The latter being less
efficient than square matrix operations.


\section{Toward an optimal algorithm}

As mentioned in the introduction, the best known algebraic time
complexity for the computation of the characteristic polynomial is not
optimal in the sense that it is not $\GO(n^\omega)$ but $\GO(n^\omega\text{log}(n))$. However,
Keller-Gehrig gives a third algorithm (let us name it \texttt{KG3}),
having this time complexity but only working on generic matrices.

To get rid of the extra $\text{log}(n)$ factor, it is no longer based on a
Krylov approach. The algorithm is inspired by a $\GO(n^3)$ algorithm by
Danilevski (described in \cite{HouseHolder:1964:TOMINA}), improved
into a block algorithm. The genericity assumption ensures the
existence of a series of similarity transformations changing the input
matrix into a companion matrix.

\subsection{Comparing the constants}

The optimal ``big-O'' complexity often hides a large constant in the exact
expression of the time complexity. This makes these algorithms 
impracticable since the improvement induced is only significant for huge
matrices. However, we show in the following lemma that
the constant of \texttt{KG3} has the same magnitude as the one of \texttt{LUK}.

\begin{lem}\label{LEM:CTE}
The computation of the characteristic polynomial of a $n \times n$
generic matrix using \texttt{KG3} algorithm requires $ K_\omega n^\omega +
o(n^\omega)$ algebraic operations, where
\begin{eqnarray*}
K_\omega & = &
C_\omega \left[-\frac{2^{\omega-2}}{2(2^{\omega-2}-1)(2^{\omega-1}-1)(2^\omega-1)}
  -\frac{1}{2^\omega-1} \right.\\
& & +\frac{1}{(2^{\omega-2}-1)(2^{\omega-1}-1)} - \frac{3}{2^{\omega-1}-1} +\frac{2}{2^{\omega-2}-1}\\
& &\left. +\frac{1}{(2^{\omega-2}-1)(2^\omega-1)}+\frac{2^{\omega-2}}{2(2^{\omega-2}-1)(2^{\omega-1}-1)^2}\right]
\end{eqnarray*}
and $C_\omega$ is the constant in the algebraic time  complexity of the matrix multiplication.
\end{lem}

The proof and a description of the algorithm are given in appendix
\ref{app:cte}.

In particular, when using classical matrix arithmetic ($\omega=3,
C_\omega=2$), we have on the one hand $K_\omega = 176/63 \approx 2.794$.

On the other hand, the algorithm \ref{alg:luk} called on a generic matrix
simply computes the $n$ Krylov vectors $A^iv$ ($2n^3$ operations), computes
the LUP factorization of these vectors ($2/3n^3$ operations) and the
coefficients of the polynomial by the resolution of a triangular
system ($\GO(n^2)$).
Therefore, the constant for this algorithm is $2+2/3 \approx 2.667$. 
These
two algorithms have thus a similar algebraic complexity, LU-Krylov being
slightly faster than Keller-Gehrig's third algorithm. We  now
compare them in practice.

\subsection{Experimental comparison}

We claim that the study  of precise algebraic time complexity of these
algorithms is worth-full in practice. Indeed these estimates directly
correspond to the computational time of these algorithms applied over
finite fields.
Therefore we ran these algorithms on a small prime finite field (word
size elements with modular arithmetic). Again we used {\tt
  modular<double>}
and {\tt FFLAS-FFPACK}. These routines can
use fast matrix arithmetic, we, however, only used classical matrix
multiplication
so as to compare two $\GO(n^3)$ algorithms having similar constants
($2.67$ for {\tt LUK} and $2.794$ for \texttt{KG3}).
We used random dense matrices over the finite field
$\mathbb{Z}_{65521}$, as generic matrices. We report the computational
speed in Mfops (Millions of field operations per second) for the
two algorithms on figure \ref{fig:lukvskgf}: 
\begin{figure}[htbp] 
\includegraphics[width=\textwidth]{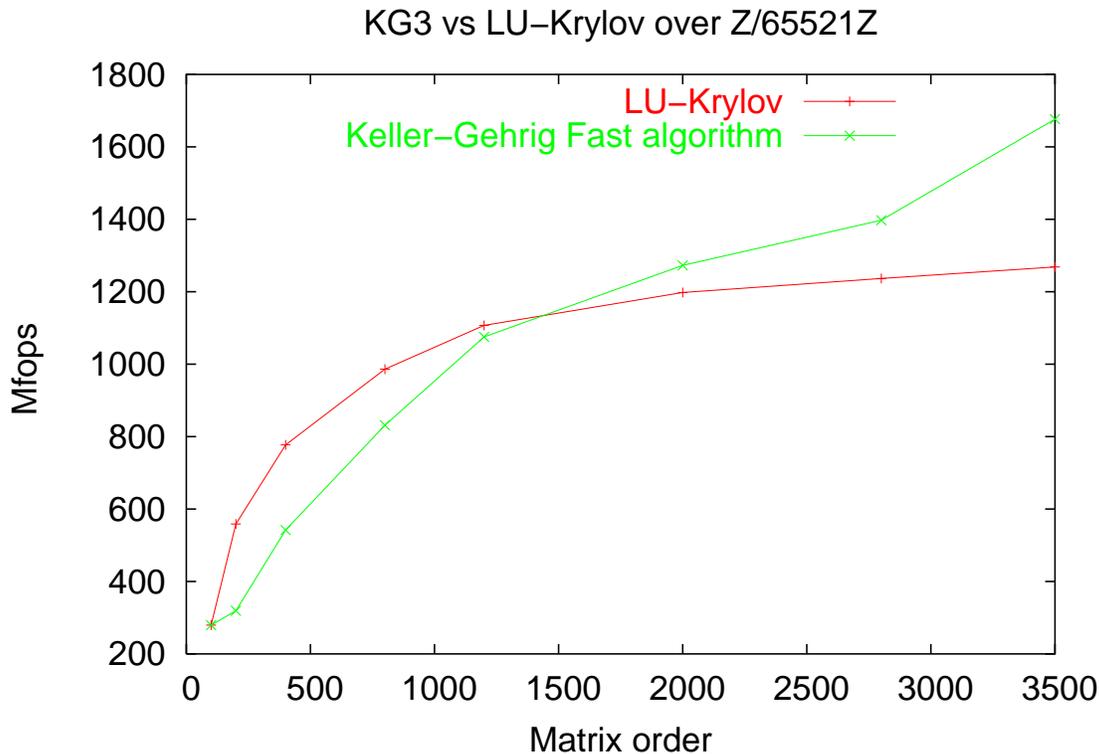} 
\caption{LUK vs. KG3: speed comparison} \label{fig:lukvskgf}
\end{figure}


It appears that LU-Krylov is faster than \texttt{KG3} for small
matrices, but for matrices of order larger than 1500, \texttt{ KG3} is
faster.
Indeed, the $\GO(n^3)$ operations
are differently performed: LU-Krylov computes the Krylov basis by $n$
matrix-vector products, whereas \texttt{KG3} only uses matrix
multiplications. Now, as the order of the matrices gets larger, the
BLAS routines provides better efficiency for matrix multiplications than
for matrix vector products. 
Once again, algorithms exclusively based on matrix multiplications
 are preferable: from the complexity point of view, they
make it possible to achieve $\GO(n^\omega)$ time complexity. In practice, 
they promise the best efficiency thanks to the BLAS better
memory management.


\newcommand{\Z}{\ensuremath{\mathbb Z}}

\section{Over the Integers}


There exist several algorithms to compute the characteristic
polynomial of an integer matrix. A first idea is to perform the
algebraic operations over the ring of integers, using exact divisions
\cite{AbMa:2001:charpdomain} or by avoiding divisions
\cite{berkowitz:84,Kaltofen:92:det,Eberly:2000:BBFDOSF,KaltofenVillard:2004:det}.
We focus here on field approaches.
Concerning the bit complexity of this computation, a first approach,
using Chinese remaindering gives $\GO\text{\~\ }(n^{\omega+1}
\text{log} \|A\|)$ bit operations ($\GO\text{\~ \ }$ is the ``soft-O''
notation, hiding logarithmic and poly-logarithmic factors in $n$ and
$\|A\|$).
 Baby-step Giant-step techniques applied by Eberly \cite{Eberly:2000:BBFDOSF} improves this
complexity to $\GO\text{\~\ }(n^{3.5} \text{log} \|A\|)$ (using
classic matrix arithmetic). 
Lastly, the recent improvement of \cite[\S
  7.2]{KaltofenVillard:2004:det}, combining Coppersmith's
block-Wiedemann techniques 
\cite{Coppersmith:1994:SHL,Kaltofen:1995:ACB,Gilles:study,Gilles:further}
set the best known exponent for this computation to $2.697263$ using fast
matrix arithmetic.


Our goal here is not to give an exhaustive comparison of these
methods, but to show that a straightforward application of our finite
field algorithm \texttt{LU-Krylov} is already very efficient and can
outperform the best existing softwares.

A first deterministic algorithm, using Chinese remaindering is given
in section \ref{ssec:dd}. Then we improve it in section \ref{ssec:early}
into a probabilistic algorithm by using the early termination
technique of \cite[\S 3.3]{DumasSaundersVillard:2001:JSC}. Therefore,
the minimal number of homomorphic computations is achieved. 
Now, for the sparse case, a recent
alternative \cite{Storjohann:2000:Frob}, also developed in \cite[\S
7.2]{KaltofenVillard:2004:det}, change the Chinese remaindering by a
Hensel p-adic lifting in order to improve the binary complexity of
the algorithm. In section \ref{ssec:hensel}, we combine some of these ideas with the Sparse Integer
  Minimal Polynomial computation of  \cite{DumasSaundersVillard:2001:JSC} and our dense modular
  characteristic polynomial to present an efficient  practical implementation.

\subsection{Dense deterministic~: Chinese remaindering}\label{ssec:dd}
The first naive way of computing the characteristic polynomial is to
use Hadamard's bound \cite[Theorem 16.6]{Gathen-Gerhard:1999}
to show that any integer coefficient of the characteristic polynomial
has the order of $n$ bits:

\begin{lem}\label{lem:hadamard}
 Let $A \in \Z^{n \times n}$, with $n\geq4$, whose coefficients are bounded in absolute
value by $B>1$. The coefficients of the characteristic polynomial
of $A$
have less than $\lceil \frac{n}{2}\left(\log_2(n)+\log_2(B^2)+1.6669\right)
\rceil$ bits.
\end{lem}
\begin{proof}
$c_i$, the $i$-th coefficient of the characteristic polynomial, is a
sum of all the $(n-i) \times (n-i)$ diagonal minors of $A$. It is therefore bounded
by ${n \choose i} \sqrt{(n-i)B^2}^{(n-i)}$. 
The lemma is true for $i=n$ since the characteristic
polynomial is unitary and also true for $i=0$ by Hadamard's bound.
Now, using Stirling's
formula ($n!<(1+\epsilon)\sqrt{2\pi n}\frac{n^n}{e^n}$), one gets
${n \choose i} <
\frac{1+\epsilon}{\sqrt{2\pi}}\sqrt{\frac{n}{i(n-i)}}\left(\frac{n}{i}\right)^i\left(\frac{n}{n-i}\right)^{n-i}$.
Thus $\log_2(c_i) < \frac{n}{2}\left(\log_2(n)+\log_2(B^2)+C\right) - H$.
Well, suppose on the first hand that $i\leq\frac{n}{2}$, 
then $H \sim (\frac{1}{ln(2)}-1+C)\frac{n}{2}+\frac{3i^2}{n ln(2)} +\frac{C-3}{2}i-\frac{5i}{2ln(2)} $.
On the
second hand, if $(n-i)\leq\frac{n}{2}$,
then
$H \sim  (\frac{1}{ln(2)}-1)n+\frac{3k^2}{n
  ln(2)}+\frac{C+5}{2}k-\frac{7k}{2ln(2)}$, where $k=n-i$. Both
equivalences are positive as soon as $C<1.6668979201$.
\end{proof}

For example, the characteristic polynomial of
$$\left[ \begin{array}{ccccc}
1&1&1&1&1\\
1&1&-1&-1&-1\\
1&-1&1&-1&-1\\
1&-1&-1&1&-1\\
1&-1&-1&-1&1
\end{array} \right]$$
is $X^5-5X^4+40X^2-80X+48$ and $80 = {5 \choose 1} \sqrt{4}^4$ is
greater than Hadamard's bound $56$, but less than our bound $1004.4$.

Note that this bound improves the one used in
\cite[lemma 2.1]{GiesbrechtStorj:2002:intratform} since $1.6669 <
2+\text{log}_2(e)\approx 3.4427$.

Now, using fast integer arithmetic and the fast 
Chinese remaindering algorithm \cite[Theorem
10.25]{Gathen-Gerhard:1999}, one gets the overall complexity for the
dense 
integer characteristic polynomial via Chinese remaindering 
of $$O(n^4 (log(n) + log(B))).$$ 
Well, as we see in next section, to go faster, the idea is actually 
to stop the remaindering earlier. Indeed, the actual coefficients can
be much smaller than the bound of lemma \ref{lem:hadamard}.

\subsection{Dense probabilistic Monte-Carlo~: early termination}\label{ssec:early}

We just use the early termination of 
\cite[\S 3.3]{DumasSaundersVillard:2001:JSC}. There it is used to stop
the remaindering of the integer minimal polynomial, here we use it to
stop the remaindering of the characteristic polynomial:
\begin{lem}\cite{DumasSaundersVillard:2001:JSC} 
Let $v \in \Z$ be a coefficient of the characteristic polynomial,
and $U$ be a given upper bound on $|v|$.
Let $P$ be a set of primes and
let $\lbrace p_1 \ldots p_k, p^{*} \rbrace$ be a random subset of $P$.
Let $l$ be a lower bound such that $p^{*} > l$ and
let $M = \prod_{i=1}^k p_i$. Let $v_k = v \mod M$,
$v^{*} = v \mod p^{*}$ and $v_k^{*} = v_k \mod p^{*}$ as above.
Suppose now that $v_k^{*} = v^{*}$. Then $v = v_k$ with probability at
least $1 - \frac{ \log_l (\frac{U-v_k}{M}) }{|P|}$.
\end{lem} \label{lem:jsc}
The proof is that of \cite[lemma 3.1]{DumasSaundersVillard:2001:JSC}.
The probabilistic algorithm is then straightforward: after each
modular computation of a characteristic polynomial, the algorithm
stops if every coefficient is unchanged. It is of the Monte-Carlo
type: always fast with a controlled probability of success.
The probability of success is bounded by the probability of lemma \ref{lem:jsc}. In practice this
probability is much higher, since the $n$ coefficients are checked.
But since they are not independent, we are not able to produce a tighter bound.

\subsection{Experimental results}

We implemented these two methods using 
\texttt{LU-Krylov}  over finite fields as described in section
\ref{sec:krylovexp}. 
The choice of  moduli is there linked to the constraints
of the matrix multiplication of \texttt{FFLAS}. Indeed, the wrapping
of numerical BLAS matrix multiplication is only valid if
$n(p-1)^2<2^{53}$ (the result can be stored in the $53$ bits of the
\texttt{double} mantissa). Therefore, we chose to sample the primes 
between $2^m$ and $2^{m+1}$ (where $m=\lfloor
25.5-\frac{1}{2}\text{log}_2(n)\rfloor$). This set was always
sufficient in practice. 
Even
with $5000\times 5000$ matrices, $m=19$ and there are $38658$ primes
between $2^{19}$ and $2^{20}$. Now if the coefficients of the matrix are
between $-1000$ and $1000$, the upper bound on the coefficients of the
characteristic polynomial is $\text{log}_{2^m}(U)\approx 4458.7$. Therefore,
the probability of finding a bad prime is lower than $4458.7/38658
\approx 0.1153$. Then performing a couple a additional modular
computations to check the result will improve this probability. 
In
this example, only 17 more computations 
(compared to the $4459$ required for the
deterministic computation)  are enough to ensure a
probability of error lower than $2^{-50}$, 
for which Knuth \cite[\S 4.5.4]{Knuth:1997:SA} considers
that there is more chances that cosmic radiations perturbed the output!

In the following, we denote by \texttt{ILUK-det} the deterministic
algorithm of section \ref{ssec:dd}, by \texttt{ILUK-prob} the
probabilistic algorithm of section \ref{ssec:early} with primes chosen
as above and by \texttt{ILUK-QD} the quasi-deterministic algorithm obtained by applying
\texttt{ILUK-prob} plus a sufficient number of modular computations to
ensure a probability of failure lower than $2^{-50}$. 

\begin{table}[htbp]
\begin{small}
\begin{center}
\begin{tabular}{|r||r|r|r|r|r|}
\hline
\textbf{$n$}  &{\tt Maple} &{\tt Magma}   & {\tt ILUK-det} & {\tt ILUK-prob}&\texttt{ILUK-QD}\\
\hline
\hline
100  & 163s   &  0.34s          & 0.22s  & 0.17s & 0.2s\\
\hline
200  & 3355s  &  4.45s          & 4.42s          & 3.17s&3.45s \\ 
     &        & 11.1Mb&3.5Mb &3.5Mb &3.5Mb \\
\hline
400  & 74970s &  69.8s          & 91.87s            & 64.3s&66.75s  \\  
     &        &56Mb&10.1Mb& 10.1Mb& 10.1Mb \\
\hline
800  &        &  1546s          & 1458s & 1053s &1062s\\ 
     &        &  403Mb          &36.3Mb&36.3Mb&36.3Mb \\
\hline
1200 &       &  8851s   & 7576s  & 5454s& 5548s  \\  
     &       &  1368Mb  & 81Mb    &  81Mb& 81Mb\\
\hline
1500 &       &   MT    & 21082s& 15277s & 15436s\\
     &       &         &  136Mb & 136Mb   & 136Mb\\
\hline
2000 &       &   MT    & 66847s & 46928s  &\\
     &       &         &  227Mb & 227Mb& \\
\hline
2500 &       &   MT    & 169355s& 124505s &\\
     &       &         &  371Mb &  371Mb &\\
\hline
3000 &       &   MT    & 349494s& 254358s&\\
     &       &       &  521Mb &  521Mb&\\
\hline
\end{tabular}
\end{center}
\end{small}
\caption{Characteristic polynomial of a dense integer matrix of order $n$ (computation
  time in seconds and memory allocation in Mb) }
\label{table:intdensecharp}
\end{table}
We report in table \ref{table:intdensecharp} the timings of their
implementations, compared to the timings of the same computation
using \texttt{Maple-v9} and \texttt{Magma-2.11}. We ran 
these tests on an athlon 2200 (1.8 Ghz) with 2Gb of RAM, running
Linux-2.4.\footnote{We are grateful to the Medicis computing center hosted by
  the CNRS STIX laboratory~: \url{http://medicis.polytechnique.fr/medicis/}.}
The matrices are formed by integers chosen uniformly between 0 and 10:
  therefore, their minimal polynomial equals their characteristic polynomial.

The implementation of Berkowitz algorithm used by \texttt{Maple} has
prohibitive computational timings. \texttt{Magma} is much faster thanks
to a $p\_$adic algorithm (probabilistic ?)
\footnote{\url{http://www.msri.org/info/computing/docs/magma/text751.htm}}.
However, no literature exists to our knowledge, describing this algorithm.
Our deterministic algorithm has similar computational timings 
and gets faster for large matrices. For matrices of
order over $800$, \texttt{magma} tries to allocate more than 2Gb of
RAM, and the computation crashes (denoted by MT as Memory Thrashing).
The memory usage of our implementations is much smaller than in
\texttt{magma}, and makes it possible to handle larger matrices.

The probabilistic algorithm \texttt{ILUK-prob} improves the computational time of
the deterministic one of roughly $27$ \%, and the cost of the extra
checks done by \texttt{ILUK-QD} is negligible.

However, this approach does not take advantage of the structure of the
matrix nor of the degree of the minimal polynomial, as 
\texttt{magma} seems to do. In the following, we will describe a
third approach to fill this gap.

\subsection{Structured or Sparse probabilistic
 Monte-Carlo}\label{ssec:hensel}
By structured or sparse matrices we mean matrices for which the
matrix-vector product can be performed with less than $n^2$ arithmetic
operations or matrices having a small minimal polynomial degree.
In those cases our idea is to compute first the integer minimal
polynomial via the specialized methods of \cite[\S
3]{DumasSaundersVillard:2001:JSC} (denoted by \texttt{IMP}), to factor it and them to simply
recover the factor exponents by a modular computation of the
characteristic polynomial.
The overall complexity is not better than
e.g. \cite{Storjohann:2000:Frob,KaltofenVillard:2004:det} but the practical
speeds shown on table \ref{table:intstructcharpoly} speak for
themselves. The algorithm is as follows:
\begin{algorithm}
\caption{\texttt{CIA}~: Characteristic polynomial over Integers Algorithm}
\label{alg:CIA}
\begin{algorithmic} [1]
\REQUIRE $A \in \Z^{n \times n}$, even as a blackbox, $\epsilon$.
\ENSURE The characteristic polynomial of $A$ with a probability of $1-\epsilon$.
\STATE $\eta = 1-\sqrt{1-\epsilon}$
\STATE $P_\text{min}^A$ = \texttt{IMP}($A,\eta$) via
\cite[\S 3]{DumasSaundersVillard:2001:JSC}.
\STATE Factor $P_\text{min}^A$ over the integers, e.g. by Hensel's lifting.
\STATE $B=2^{ \frac{n}{2}\left(\log_2(n)+\log_2(||A||^2)+1.6669\right)}$
\STATE Choose a random prime $p$ in a set of
$\frac{1}{\eta}\text{log}_2(\sqrt{n+1}\ 2^{n+1}B+1)$ primes. 
\STATE Compute  $P_p$ the characteristic polynomial of $A$ mod $p$ via \texttt{LUK}.
\FORALL{$f_i$ irreducible factor of $P_\text{min}^A$}
\STATE Compute $\overline{f}_{i} \equiv f_i \mod p$.
\STATE Find $\alpha_i$ the multiplicity of $\overline{f}_{i}$ within $P_p$.
\IF{$\alpha_i$ == 0}
\STATE Return ``FAIL''.
\ENDIF
\ENDFOR
\STATE Compute $P_\text{char}^A = \prod f_i^{\alpha_i}
=X^n-\sum_{i=0}^{n-1}{a_iX^i}$.
\IF{ ($\sum \alpha_i \text{degree}(f_i) \neq n$)  }
\STATE Return ``FAIL''.
\ENDIF
\IF{ ($Trace(A) \neq a_{n-1}$ )}
\STATE Return ``FAIL''.
\ENDIF
\STATE Return $P_\text{char}^A$.
\end{algorithmic}
\end{algorithm}
\begin{thm} 
Algorithm \ref{alg:CIA} is correct. It is probabilistic of the
Monte-Carlo type. 
Moreover, most cases where the result is wrong are identified.
\end{thm}
\begin{proof}
Let $P^\text{min}$ be the integer minimal polynomial of $A$ and
$\tilde P^\text{min}$ the result of the call to \texttt{IMP}.

With a probability of $\sqrt{1-\epsilon} $, $P^\text{min}=\tilde
P^\text{min}$. Then the only problem that can occur is that an irreducible factor of
  $P^\text{min}$ divides another factor when taken modulo $p$, or
equivalently, that $p$ divides the resultant of these polynomials. Now
from \cite[Algorithm 6.38]{Gathen-Gerhard:1999} and lemma
\ref{lem:hadamard} an upper bound on the size of this 
resultant is $\text{log}_2(\sqrt{n+1}\ 2^{n+1}B+1)$. Therefore, the
probability of choosing a bad prime is less than $\eta$.
Thus the result will be correct with a probability greater than
$1-\epsilon$
\end{proof}

This algorithm is also able to detect most erroneous results and return
``FAIL'' instead. We call it therefore ``Quasi-Las-Vegas''.

The first case is when $P^\text{min}=\tilde P^\text{min}$ and a factor of
$P^\text{min}$ divides another factor modulo $p$.
In such a case, the exponent of this factor will appear twice in
the reconstructed characteristic polynomial. The overall degree being
greater than $n$, \texttt{FAIL} will be returned. 

Now, if $P^\text{min}\neq \tilde P^\text{min}$, the tests $\alpha_i>0$ will
detect it unless $\tilde P^\text{min}$ is a divisor of
$P^\text{min}$, say $P^\text{min} = \tilde P^\text{min} Q$. In that 
case, on the one hand, if $Q$ does not divide  $\tilde P^\text{min}$ modulo $p$, the
total degree will be lower than $n$ and \texttt{FAIL} will be
returned. On the other hand, a wrong characteristic polynomial will be
reconstructed, but the trace test will detect most of these cases.


We now compare our algorithms to \texttt{magma}.
In  table \ref{table:intstructcharpoly}, we denote by $d$ the degree of the integer
minimal polynomial and by $\omega$ the average number of nonzero
elements per row within the sparse matrix. \texttt{CIA} is written in C++ and
uses different external modules:
the integer minimal polynomial is
computed with LinBox\footnote{\url{www.linalg.org}} via \cite[\S
3]{DumasSaundersVillard:2001:JSC}, the polynomial factorization is
computed with NTL\footnote{\url{www.shoup.net/ntl}} via Hensel's
factorization.

\begin{table}[htbp]
\begin{scriptsize}
\begin{center}
\begin{tabular}{|c||r|r|r|r|r|r|}
\hline
Matrix           &  $A$ &$U^{-1}AU$&$A^TA$ & B
&$U^{-1}BU$& $B^TB$  \\
\hline
$n$              & 300  & 300  & 300   & 600  & 600 & 600 \\
$d$              & 75   & 75   & 21    & 424  & 424  & 8\\
$\omega$         & 1.9  & 300  & 2.95  & 4    & 600 & 13\\
\hline
\texttt{ILUK-prob}& 1.3 & \textbf{1.5} & 18.3& 31.8 & \textbf{34.9} & 120.0\\
\hline
\texttt{ILUK-det}& 37.5 & 121.7 & 265.0 & 310 &3412  &422.3 \\
\hline
\texttt{Magma}   & 1.4 & 16.5& \textbf{0.2}  & 6.2 & 184.0 & 6.0 \\
\hline
\texttt{CIA}    & \textbf{0.32} & 3.72 & 0.86 & \textbf{4.51} & 325.1 &\textbf{2.4} \\
\multicolumn{1}{|r||}{\texttt{IMP}}
 & 0.01 & 3.38 & 0.01& 1.49  & 322.1  & 0.04\\
\multicolumn{1}{|r||}{\texttt{Fact} }
   & 0.05 & 0.05 & 0.01& 0.76 & 0.76  & 0.01\\
\multicolumn{1}{|r||}{\texttt{LUK}+\texttt{Mul}}
     & 0.26 & 0.29 & 0.84& 2.26 & 2.26 & 2.30\\
\hline
\end{tabular}\end{center}
\caption{\texttt{CIA} on sparse or structured matrices}\label{table:intstructcharpoly}
\end{scriptsize}
\end{table}


We show the computational times of algorithm \ref{alg:CIA}
(\texttt{CIA}), decomposed into the time for the integer minimal
polynomial computation (\texttt{IMP}), the factorization of this
polynomial (\texttt{Fact}), the computation of the characteristic
polynomial and  the computation of the multiplicities
(\texttt{LUK}+\texttt{Mul}). They are compared to the timings of the
algorithms of section \ref{ssec:dd} and \ref{ssec:early}.

We used two sparse matrices $A$ and $B$ of order $300$ and $600$,
having a minimal polynomial of degree respectively $75$ and $424$. $A$
is the almost empty  matrix
\texttt{Frob08blocks} and is in Frobenius normal form
with $8$ companion blocks and $B$ is the matrix
\texttt{ch5-5.b3.600x600\-.sms} presented in
\cite{DumasSaundersVillard:2001:JSC}.

On these matrices \texttt{magma} is pretty efficient thanks to their
sparsity. The early termination in \texttt{ILUK-prob} gives similar
timings for $A$, since the coefficients of its characteristic
polynomial are small. But this is not the case with $B$.
\texttt{ILUK-det} performs many useless operations since the Hadamard
bound is well overestimating the size of the coefficients. 
\texttt{CIA} also takes advantage of both the sparsity
and the low degree of the minimal polynomial. It is actually much
faster than \texttt{magma} for $A$ and is slightly faster for $B$ (the
degree of the minimal polynomial is bigger).

Then, we made these matrices dense with an integral similarity
transformation. The lack of sparsity slows down both \texttt{magma}
and \texttt{CIA}, whereas \texttt{ILUK-prob}
maintains similar timings. \texttt{ILUK-det} is much
slower because the bigger size of the matrix entries increases the Hadamard
bound. 

Lastly, we used symmetric matrices with small minimal polynomial
($A^TA$ and $B^TB$). The bigger size of the coefficients of the
characteristic polynomial makes the Chinese remainder methods of
\texttt{ILUK-prob} and \texttt{ILUK-det} slower. \texttt{CIA} is still
pretty efficient ( the best on $B^TB$ ), but \texttt{magma} appears to be
extremely fast on $A^TA$.

We report in table \ref{table:iscp2} on some comparisons using other sparse
matrices\footnote{These matrices are available at \url{http://www-lmc.imag.fr/lmc-mosaic/Jean-Guillaume.Dumas/Matrices}}.
\begin{table}[htbp]
\begin{small}
\begin{center}
\begin{tabular}{|c||r|r|r|r|r|}
\hline
Matrix           &  $n$ & $\omega$ & \texttt{magma} & \texttt{CIA} & \texttt{ILUK-QD}  \\
\hline
\texttt{TF12}    & 552  & 7.6  & 10.03s   & 6.93s  & 51.84s  \\
\texttt{Tref500} & 500  & 16.9 & 108.1s   & 64.58s & 335.04s \\
\texttt{mk9b3}   & 1260 &  3   & 77.02s   & 35.74s & 348.31s \\
\hline
\end{tabular}\end{center}
\caption{\texttt{CIA} on other sparse matrices}\label{table:iscp2}
\end{small}
\end{table}


To conclude, \texttt{ILUK-det} is always too expensive, although it has
better timings than \texttt{magma} for large dense matrices (cf. table
\ref{table:intdensecharp}). \texttt{ILUK-prob} is well suited for
every kind of matrix having a characteristic polynomials with small
coefficients. Now with sparse or structured matrices, \texttt{magma}
and \texttt{CIA} are more efficient; \texttt{CIA} being almost
always faster.


\section{Conclusion}

We presented a new algorithm for the computation of the characteristic
polynomial over a finite field, and proved its efficiency in practice.
We also considered Keller-Gehrig's third algorithm and showed that its
generalization would be not only interesting in theory but produce a
practicable algorithm.

We applied our algorithm for the computation of the integer
characteristic polynomial in two ways: a combination of Chinese
remaindering and early termination for dense matrix computations, and
a mixed blackbox-dense algorithm for sparse or structured matrices.
These two algorithm outperform the existing software for this task.
Moreover we showed that the recent improvements of
\cite{Storjohann:2000:Frob,KaltofenVillard:2004:det} should be highly
practicable since the successful \texttt{CIA} algorithm is inspired
by their ideas. It remains to show how much they improve the
simple approach of \texttt{CIA}.

To improve the dense matrix computation over a finite field, one should
consider the generalization of Keller-Gehrig's third algorithm. 
At least some heuristics could be built: using row-reduced form
elimination to give produce generic rank profile.

Lastly, concerning the sparse computations, the blackbox algorithms of
\cite{Villard:2000:Frob} and of \cite{Eberly:2004:krylov}, could
handle huge sparse matrices (no dense computation is used as in
\texttt{CIA}). But one should study how their use of 
preconditionners, expensive in practice, penalize them.

{\footnotesize
\bibliographystyle{abbrv}
\bibliography{charpoly}  
}
\appendix
\section{On  Keller-Gehrig's third algorithm} \label{app:cte}

We first recall the principle of this algorithm, so as to determine
the exact constant in its algebraic time complexity. This advocates
for its practicability.

\subsection{Principle of the algo\-ri\-thm}

First, let us define a $m$-Frobenius form as a $n \times n$ matrix of the shape:
$\left[
\begin{matrix}
0 & M_1 \\
Id_{n - m} & M_2 
\end{matrix}
\right]
$.

Note that a $1$-Frobenius form is a companion
matrix, which characteristic polynomial is given by the opposites of
the coefficients of its last column.

The aim of the algorithm is to compute the $1$-Frobenius form $A_0$ of $A$
by computing the sequence of matrices $A_r = A, \ldots, A_0,
$ where $A_i$ has the $2^i$-Froebenius form and $r = \left\lceil
\log n \right\rceil$. The idea is to compute $A_i $ from $A_{i + 1}$ by
slicing the block $M$ of $A_{i + 1}$ into two $n \times 2^i$ columns
blocks $B$ and $C$. Then, similarity transformations with the matrix
\[ U = 
\left[
\begin{matrix}
0 & C_1\\
Id_{n - 2^i} & C_2
\end{matrix}
\right]
\]
will ``shift'' the block $B$ to the left and generate an identity
block of size $2^i$ between $B$ and $C$.

\begin{figure}[htpb]
\begin{center}
\includegraphics[width=.3\textwidth]{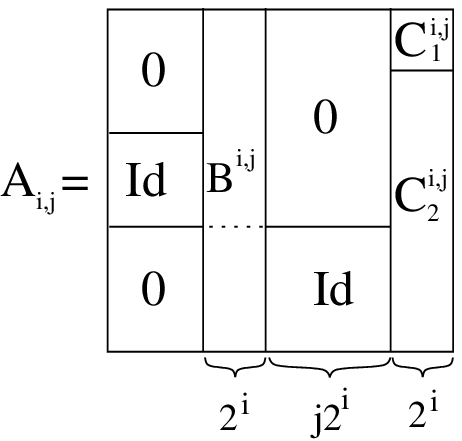}
\includegraphics[width=.3\textwidth]{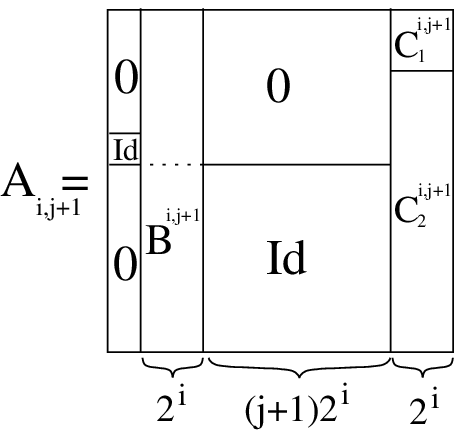}
\includegraphics[width=.3\textwidth]{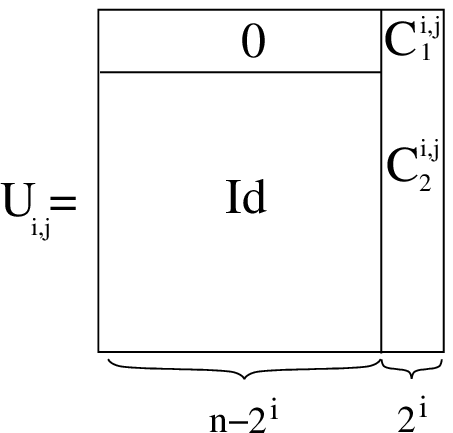}
\end{center}
\caption{Principle of Keller-Gehrig's third algorithm}
\label{fig:kgf}
\end{figure}

More precisely, the algorithm computes the sequence of matrices 
$A_{i, 0} = A_{i + 1}, A_{i, 1}, \ldots, A_{i, s_i} = A_i$, 
where 
$s_{i^{}} = \left\lceil n/ 2^i \right\rceil - 1$, by the relation
$A_{i, j + 1} = U_{i, j}^{- 1} A_{i,j} U_{i, j}$, whith the notations
of figure \ref{fig:kgf}.

As long as $C_1$ is invertible, the process will carry on, and make at
last the block $B$ disapear from the matrix. This last condition is
restricting and is the reason why this algortihm is only valid for
generic matrices.

\subsection{Proof of lemma \ref{LEM:CTE}}

\textsc{Lemma \ref{LEM:CTE}.}
\textit{
The computation of the characteristic polynomial of a $n \times n$
generic matrix using the fast algorithm requires $ K_\omega n^\omega +
o(n^\omega)$ algebraic operations, where
\begin{eqnarray*}
K_\omega & = &
C_\omega \left[-\frac{2^{\omega-2}}{2(2^{\omega-2}-1)(2^{\omega-1}-1)(2^\omega-1)}
  -\frac{1}{2^\omega-1} \right.\\
& & +\frac{1}{(2^{\omega-2}-1)(2^{\omega-1}-1)} - \frac{3}{2^{\omega-1}-1} +\frac{2}{2^{\omega-2}-1}\\
& &\left. +\frac{1}{(2^{\omega-2}-1)(2^\omega-1)}+\frac{2^{\omega-2}}{2(2^{\omega-2}-1)(2^{\omega-1}-1)^2}\right]
\end{eqnarray*}
and $C_\omega$ is the constant in the algebraic time  complexity of the matrix multiplication.
}
\begin{proof}
We will denote by $X_{a \dots b}$ the submatrix composed by the
rows from $a$ to $b$ of the block $X$.
For a given $i$, \texttt{KG3}  performs $n/2^i$ similarity
transformations. Each one of them can be described by the following operations:

\begin{algorithmic} [1]
\STATE $B'_{n-2^i+1 \dots n} = C^{-1}_{1 \dots 2^i}B_{1 \dots 2^i}$
\STATE $B'_{1 \dots n-2^i} = -C_{2^i+1 \dots n} B'_{n-2^i+1 \dots n} +
  B_{2^i+1 \dots n}$
\STATE $C' = B'C_{\lambda+1 \dots \lambda+2^i}$
\STATE $C'_{2^i+1 \dots 2^i+\lambda} += C_{1 \dots \lambda}$
\STATE $C'_{2^i+\lambda+1 \dots n} += C_{2^i+\lambda+1 \dots n}$
\end{algorithmic}

The first operation is a system resolution, and consists in a LUP
factorization and two triangular system solve with matrix right hand
side. The two following ones are matrix multiplications, and we do not
consider the two last ones, since their cost is dominated by the
previous ones. The cost of a similarity transformation is then:
\begin{eqnarray*}
T_{i,j} &=& T_{\text{LUP}}(2^i,2^i) + 2 T_{\text{TRSM}}(2^i,2^i) \\
& & + T_{\text{MM}}(n-2^i,2^i,2^i)+ T_{\text{MM}}(n,2^i,2^i)
\end{eqnarray*}

From \cite[Lemma 4.1]{Dumas:2004:FFPACK} and \cite{Pernet:2003:dea},
we have
$$
T_{\text{LUP}}(m,n) = \frac{C_\omega}{2^{\omega-1}-2}m^{\omega-1}\left(n-m\frac{2^{\omega-2}-1}{2^{\omega-1}-1}\right)
$$
and 
$$
T_{\text{TRSM}}(2^i,2^i) = \frac{C_\omega mn^{\omega-1}}{2\left(2^{\omega-1}-1\right)}
$$
Therefore
\begin{eqnarray*}
T_{i,j} &=& \frac{C_\omega
  2^{\omega-2}}{2\left(2^{\omega-2}-1\right)\left(2^{\omega-1}-1\right)}(2^i)^\omega+\frac{C_\omega}{\left(2^{\omega-2}-1\right)}(2^i)^\omega\\
& &+C_\omega(n-2^i)(2^i)^{\omega-1}+C_\omega n(2^i)^{\omega-1}\\
&=& C_\omega(2^i)^\omega
  \underbrace{\left(\frac{2^{\omega-3}+2^{\omega-1}-1}{\left(2^{\omega-2}-1\right)\left(2^{\omega-1}-1\right)}-1
  \right)}_{D_\omega}\\
& &+2nC_\omega(2^i)^{\omega-1}
\end{eqnarray*}

And so the total cost of the algorithm is 
\begin{eqnarray*}
T&=&\sum_{i=1}^{log(n/2)}{\sum_{j=1}^{n/2^i-1}{T_{i,j}}}\\
 &=&\sum_{i=1}^{log(n/2)}{\left(\frac{n}{2^i}-1\right)C_\omega
 D_\omega (2^i)^\omega + 2nC_\omega(2^i)^{\omega-1}}\\
&=& C_\omega \sum_{i=1}^{log(n/2)} ( D_\omega -3)n(2^i)^{\omega-1}+2
 n^2(2^i)^{\omega-2}\\
& & - D_\omega (2^i)^\omega
\end{eqnarray*}
And since
$$
\sum_{i=1}^{log(n/2)}{(2^i)^x}=\frac{n^x-1}{2^x-1}=\frac{n^x}{2^x-1} + o(n^x)
$$
we get the result:
\begin{eqnarray*}
T & = & n^\omega C_\omega \left[-\frac{2^{\omega-2}}{2(2^{\omega-2}-1)(2^{\omega-1}-1)(2^\omega-1)}
  -\frac{1}{2^\omega-1} \right.\\
& & +\frac{1}{(2^{\omega-2}-1)(2^{\omega-1}-1)} - \frac{3}{2^{\omega-1}-1} +\frac{2}{2^{\omega-2}-1}\\
& &\left.
  +\frac{1}{(2^{\omega-2}-1)(2^\omega-1)}+\frac{2^{\omega-2}}{2(2^{\omega-2}-1)(2^{\omega-1}-1)^2}\right]
\end{eqnarray*}
\end{proof}

\end{document}